\documentclass[12pt]{iopart}

\usepackage{graphicx}
\begin{document}

\title[Energy dependence of the freeze out eccentricity from azimuthal HBT]{Energy
dependence of the freeze out eccentricity from the azimuthal
dependence of HBT at STAR}

\author{Christopher Anson (for the STAR Collaboration)}

\address{Department of Physics, The Ohio State University, Columbus, OH 43210, USA}
\ead{anson.9@osu.edu}
\begin{abstract}

     Non-central heavy ion collisions create an out-of-plane-extended participant
zone that expands toward a more round state as the system evolves.  The recent RHIC
Beam Energy Scan at $\sqrt{s_{NN}}$ of 7.7, 11.5, and 39 GeV provide an opportunity
to explore the energy dependence of the freeze out eccentricity.  The new low energy
data from STAR complements high statistics data sets at $\sqrt{s_{NN}}$ of 62.4
and 200 GeV.  Hanbury-Brown-Twiss (HBT) interferometry allows to determine the size 
of pion emitting source regions. The dependence of the HBT radius parameters on 
azimuthal angle relative to the reaction plane have been extracted.  These dependencies can be related to 
the freeze out eccentricity.  The new results from STAR are consistent with a monotonically
decreasing freeze out eccentricity and constrain any minimum, suggested by previously available data,
to lie in the range between 11.5 and 39 GeV.  Of several models UrQMD appears to 
best predict the STAR and AGS data.

\end{abstract}



\section{Introduction}

     One of the primary goals of heavy ion physics is to explore the phase diagram of QCD
matter by mapping out the nature of the transition between a deconfined and hadronic
state.  The recent Beam Energy Scan at RHIC allows to search for signs of a change in
the Equation of State of hot, dense QCD matter produced in heavy ion collisions.  One of
the main observables that is sensitive to the Equation of State is the freeze-out shape 
of the participant zone in non-central collisions. 

     At its creation, the participant zone forms an out-of-plane extended ellipsoid, with
an initial eccentricity (in the transverse plane) in non-central heavy ion collisions.  The 
material, being more compressed in the reaction plane, experiences larger in-plane pressure
gradients compared to the out-of plane direction.  Preferential in-plane 
expansion drives the participant zone toward a more circular freeze out shape.  Systems with
a longer lifetime, larger in-plane versus out-of-plane pressure gradients, or 
both, achieve a more round shape or could even concievably become in-plane extended.  Based 
solely on these two considerations, we would expect the excitation function for the freeze out 
eccentricity to fall monotonically with increasing energy.

     However, the five measurements available prior to the Beam Energy Scan at 2.7, 3.32, and 
3.84 GeV \cite{ref1}, at 17.3 GeV \cite{ref2}, and at 200 GeV 
\cite{ref3}, suggested a possible minimum followed by a rise in the excitation function with 
increasing energy.  It has been speculated \cite{ref4} that such a scenario may somehow be 
explained by a softening of the equation of state due to entrance, above some energy, into a 
mixed phase.  As the pressure gradients vanish in a mixed phase, the shape would not change 
during this portion of the lifetime.  This may be observed as a plateau or perhaps even a 
minimum in the excitation function.

     The suggestion of possible non-monotonic behaviour and the availability of BES data 
provide motivation and opportunity for extracting the freeze out shape. The means is provided
by HBT analysis relative to the event plane.

\section{Method}

	HBT analyses provide information about the shape of the particle emitting source
regions in heavy ion collisions expressed in terms of radii in the Bertsch-Pratt (out,side,long or o,s,l)
coordinate system.  In the current analysis, HBT radii are measured relative to the $2^{nd}$ order 
event plane determined using charged particles in the STAR Time Projection Chamber.  The correlation functions 
are constructed using like-signed pion pairs with average transverse pair momenta, $k_{T}=(p_{T1}+p_{T2})/2$, 
in the range $0.15<k_{T}<0.60$ GeV/c.  After corrections, discussed below, are applied to 
both the same event and mixed event pair distributions the radii are extracted by fitting
a correlation function of the form
\begin{equation}
C({\bf q}) = N\{(1-\lambda) + \lambda K_{c}[1+\exp{(-q_{o}^{2}R_{o}^{2}-q_{s}^{2}R_{s}^{2}-q_{l}^{2}R_{l}^{2}-2q_{o}q_{s}R_{os}^{2})}]\}
\end{equation}
using a fit range of $|q_{i}|<0.15$ GeV/c, $i=$ {\it o, s, l}.  N is a normalization factor, $\lambda$
the fraction of pairs that interact, $R_{i}$ the HBT radii in the out-side-long coordinate system, 
$q_{i}$ the relative momenta, ${\bf q}={\bf p_{1}}-{\bf p_{2}}$, projected on the out-side-long coordinate system,
and $K_{c}$ the Coulomb interaction for each $q_{o}$, $q_{s}$, $q_{l}$ bin.  The extracted radii exhibit
$2^{nd}$ order oscillations relative to the event plane expressed using Fourier
coefficients as

\begin{equation}
R_{i}^{2}(\phi) = R_{i,0}^{2} + 2\Sigma_{n=2,4,6,...}R_{i,n}^{2}\cos(n\phi) \qquad i={\it o,s,l}
\end{equation}
\begin{equation}
R_{i}^{2}(\phi) = 2\Sigma_{n=2,4,6,...}R_{i,n}^{2}\sin(n\phi) \qquad \qquad \quad i={\it os}
\end{equation}

\noindent where $0 < \phi < \pi$ is the angle from the event plane at which the radii are observed.

The freeze out eccentricity can be approximated \cite{ref5,ref6} using

\begin{equation}
\varepsilon_{F} = \frac{\sigma_{y}^{2}-\sigma_{x}^{2}}{\sigma_{y}^{2}+\sigma_{x}^{2}} \approx 2 \frac{R_{s,2}^{2}}{R_{s,0}^{2}},
\end{equation}

\noindent where $\sigma_{x}^{2}$ and $\sigma_{y}^{2}$ are the in- and out-of-plane length scales. More in depth 
discussion of this method of analysis can be found in \cite{ref5}.

	The amplitude of the oscillation plays a direct role in the calculation and
extracting the correct value requires correction for two effects:  finite reaction
plane resolution, and finite bin width.  Uncertainty in the computed reaction plane mixes pairs
in adjacent bins, reducing the extracted oscillation amplitude.  Similarly, pairs within a
finite bin width correspond not to single source region but to a range of source regions with 
a range of sizes.  The extracted radii in a bin are averages which again reduces the 
observed oscillation.  A model independent correction algorithm \cite{ref7} compensates for both effects.

\begin{figure}\label{fig1}
\begin{center}
\includegraphics[width=0.61\textwidth]{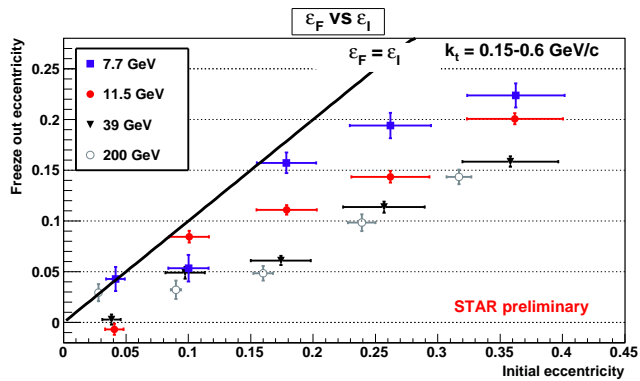} 
\end{center}
\caption{ $\varepsilon_{F}$ versus $\varepsilon_{I}$.  Points further below the line $\varepsilon_{F} = \varepsilon_{I}$ change shape more than points closer to the line.  Errors are statistical only.  Color online.}
\end{figure}

\section{Results} 

	The shape evolution for different energies is demonstrated in figure 1.  For all energies 
the freeze out eccentricity generally increases from central to peripheral collisions.  The shape 
changes relatively little for the central collisions because of the smaller difference between 
in- and out-of-plane pressure gradients.  In more peripheral collisions the shape changes
more but in all cases the evolution is such that the shape tends to become more circular while 
remaining out-of-plane extended.  The amount of change in shape, at least for the 10-80\% 
centrality range, shows an energy dependence.  The 7.7 GeV results lie closest to the line
$\varepsilon_{F}=\varepsilon_{I}$ which means the shape has changed the least.  At 11.5 GeV the
system reaches a noticeably more round shape and from 11.5 to 39 GeV there is again a noticeable 
change.  In a much wider range from 39 GeV to 200 GeV the shape continues to evolve a similar 
amount, relatively independent of collision energy.
\begin{figure}\label{fig2}
\begin{center}
\includegraphics[width=0.61\textwidth]{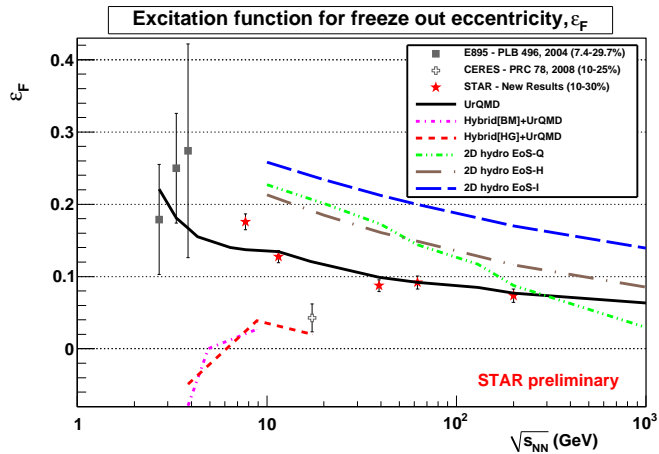} 
\end{center}
\caption{ Freeze out eccentricity, $\varepsilon_{F}$, as a function of $\sqrt{s_{NN}}$ for data and
models \cite{ref4} with similar centralities.  Color online.}
\end{figure}

	The excitation function, figure 2, is constructed by averaging the results in the
10-20\% and 20-30\% centrality range to obtain a result comparable to results from other
experiments.  The errors include statistical errors as well as preliminary systematic errors 
from track and pair cuts; further
sources of systematic error being investigated but each is expected to be small.  Taken alone,
the STAR data are consistent with a monotonically decreasing shape.  Any minimum, if confirmed,
is constrained to the region between 11.5 GeV and 39 GeV.  If a minimum is not confirmed changes in the
slope of the excitation function might still be related to changes in the equation of state 
through detailed model comparisons.

	Comparison to models, from \cite{ref4}, shows that the prediction from UrQMD comes closest to describing
the results from STAR and AGS.  The hybrid models tend to come close to the CERES point but underpredict
the rest while the 2D hydrodynamic predictions tend to overpredict the data at most energies.
An interesting comparison between 2D hydrodynamic models with hadronic (EoS-H) and ideal 
quark gas (EoS-I) equations of state shows the sensitivity of the freeze out shape
to changes in the equation of state so the new data provides an important opportunity to 
further constrain models.

\section{Summary}

	Energy dependence of the freeze out eccentricity suggest the amount of change in shape
from the initial value increases relatively rapidly at lower energies but less rapidly at higher energies.
If a minimum is confirmed it is constrained to lie between 11.5 and 39 GeV.  Recently obtained 19.6 GeV 
and 27 GeV data will allow to study the excitation function in this interesting region with the same
detector, acceptance and analysis techniques.  Detailed comparisons will further constrain models,
contributing to a more accurate description of heavy ion collisions over a range of energies 
that access much of the phase diagram of hot QCD matter.

\end{document}